# Mach-Zehnder based Rotational Shearing Interferometer for Non-destructive Testing using Spatial-Phase-Shifting Shearography

VALENTIN BASTGEN,[1,2] MICHAEL SCHUTH,[2] AND GEORG VON FREYMANN[1,3,4]

[1] *Department of Physics and Research Center OPTIMAS, RPTU University, 67663 Kaiserslautern-Landau, Germany*
[2] *Department of Engineering and Technology Center OGKB, Trier University of Applied Sciences, 54296 Trier, Germany, bastgenv@hochschule-trier.de*
[3] *Fraunhofer Institute for Industrial Mathematics ITWM, 67663 Kaiserslautern, Germany*
[4] *Institute of Photonic Systems and Technologies, Leibniz Universität Hannover, 30823 Garbsen, Germany*

**Abstract:** This work introduces a novel optical setup based on a Mach-Zehnder interferometer, enabling spatial phase-shifting shearography with rotational shear. The optical concept employs a virtual double-slit configuration, which decouples the adjustment of the shear amount from the generation of the spatial carrier frequency, thereby enabling flexible control of the measurement sensitivity. Rotational shear is generated within the optical setup by means of image rotation using Dove prisms. Since the proposed system is based on spatial phase-shifting shearography, full-field measurements can be performed at the camera frame rate, making the method suitable for fast non-destructive testing under industrial conditions. In contrast to conventional linear shear configurations, the rotational shear approach is sensitive to tangential displacement gradients around the centre of rotation. In this paper, the proposed rotational shear approach is compared with linear shear configurations, with the results demonstrating reliable detection of relevant defects. Spatial phase-shifting shearography combined with rotational shear offers significant potential for industrial non-destructive testing applications, particularly in sealing technology and for the inspection of rotationally symmetric components.

## 1. Introduction

In industry, the demand for non-destructive testing for comprehensive component monitoring is steadily increasing [1], pp. 3-4. Shearography, also referred to as electronic speckle pattern shearing interferometry (ESPSI), is an optical non-destructive testing method that operates contactless, provides full-field measurements, and is largely independent of the material under investigation. Owing to developments in spatial phase-shifting shearography (SPS), state-of-the-art systems today enable rapid measurements at the frame rate of the employed camera. The further development of the Mach-Zehnder interferometer configuration employing a virtual double slit (VDS) by C. Petry et al. [2], leads to a substantial increase in measurement light efficiency and enables free adjustment of the shear amount, decoupled from spatial carrier-frequency formation. As a result, inspections can also be performed in industrial (in-line) environments under comparatively harsh ambient conditions [3]. In industrial applications, the current state of the art in spatial phase-shifting shearography relies on linear shear configurations, typically applied along the *x*-, *y*-, or combined *xy*-directions. Conventional shearographic methods face several limitations: The usable measurement area on rotationally symmetric parts with thin walls is severely limited due to the necessity of linear shear adjustments. Multiple measurements and rotational repositioning of the object are generally required. An early setup reported in 1985 employs rotational shear in combination with a double-exposure technique [4]. However, this configuration exhibits a very low measurement light efficiency, as the aperture used at that time is positioned in front of two Dove prisms. Consequently, according to the current state of the art, industrial measurements at the camera frame rate cannot be performed. Another existing setup utilizes temporal phase shifting and is implemented as a Mach–Zehnder interferometer with a piezoelectric actuator, in which two Dove prisms serve as image rotators [5]. In comparison to spatial phase shifting, temporal phase shifting (TPS) is significantly more sensitive to vibrations during the measurement process. TPS is therefore unsuitable for industrial series inspection. To develop an in-line-capable shearography system with rotational shear, the method of spatial phase shifting is consequently employed.

The newly proposed rotational-shear setup introduces a VDS, thereby decoupling the adjusted shear amount from the formation of the spatial carrier frequency and ensuring unrestricted sensitivity adjustment. The use of SPS in combination with rotational shear enables rapid and full-field shearographic inspection of rotationally symmetric components, such as safety-critical sealing rings. Since shearography operates in a contactless manner and is largely independent of the material under investigation, its application range for non-destructive component monitoring is highly versatile.

## 2. Setup for shear-independent spatial carrier frequency formation

The fundamental optical setup based on a Mach-Zehnder interferometer with a preceding single slit aperture was introduced by G. Pedrini et al. [6]. In addition, a Michelson interferometer configuration employing spatial phase shifting with a pinhole aperture, which is placed in front of the setup has been reported [3]. When the aperture is placed in front of the interferometric setup, the resulting drawbacks include image vignetting caused by the large distance between the aperture and the camera, as well as the limited adjustability of the shear amount, which prevents flexible sensitivity adaptation in shearographic measurements. Due to diffraction of the monochromatic coherent laser light, the speckles become elongated after passing through the aperture. When a linear shear amount is applied within the optical setup (e.g., along the *x*- or *y*-direction), a coherent intensity modulation of the speckle pattern is formed [7], pp. 86 ff. By applying a shear amount orthogonal to the slit direction, a spatial carrier frequency is introduced, and the diffracted, elongated speckles are modulated. In conventional SPS configurations, the formation of this spatial carrier frequency is directly dependent on the applied shear amount and is not decoupled from it. The following equation describes the relationship between the spatial carrier frequency $f_0$, the wavelength $\lambda$ and the shear amount $\beta$ in the conventional approach [8]:

$$f_0 = \frac{\sin(\beta)}{\lambda} \tag{1}$$

The rotational shear approach introduced in this work is based on a Mach-Zehnder interferometer configuration (see fig. 1). The distinctive feature of this optical system is the arrangement of the slit apertures in each interferometer arm, which are placed in front of the "Beam Splitter 2", located close to the camera. One of the slit apertures is laterally displaced from the optical axis by the distance "a" in a direction orthogonal to the slit height. This configuration is therefore referred to in the following as a VDS [2]. The experimental SPS setup used to generate a rotational shear is schematically illustrated in figure 1.

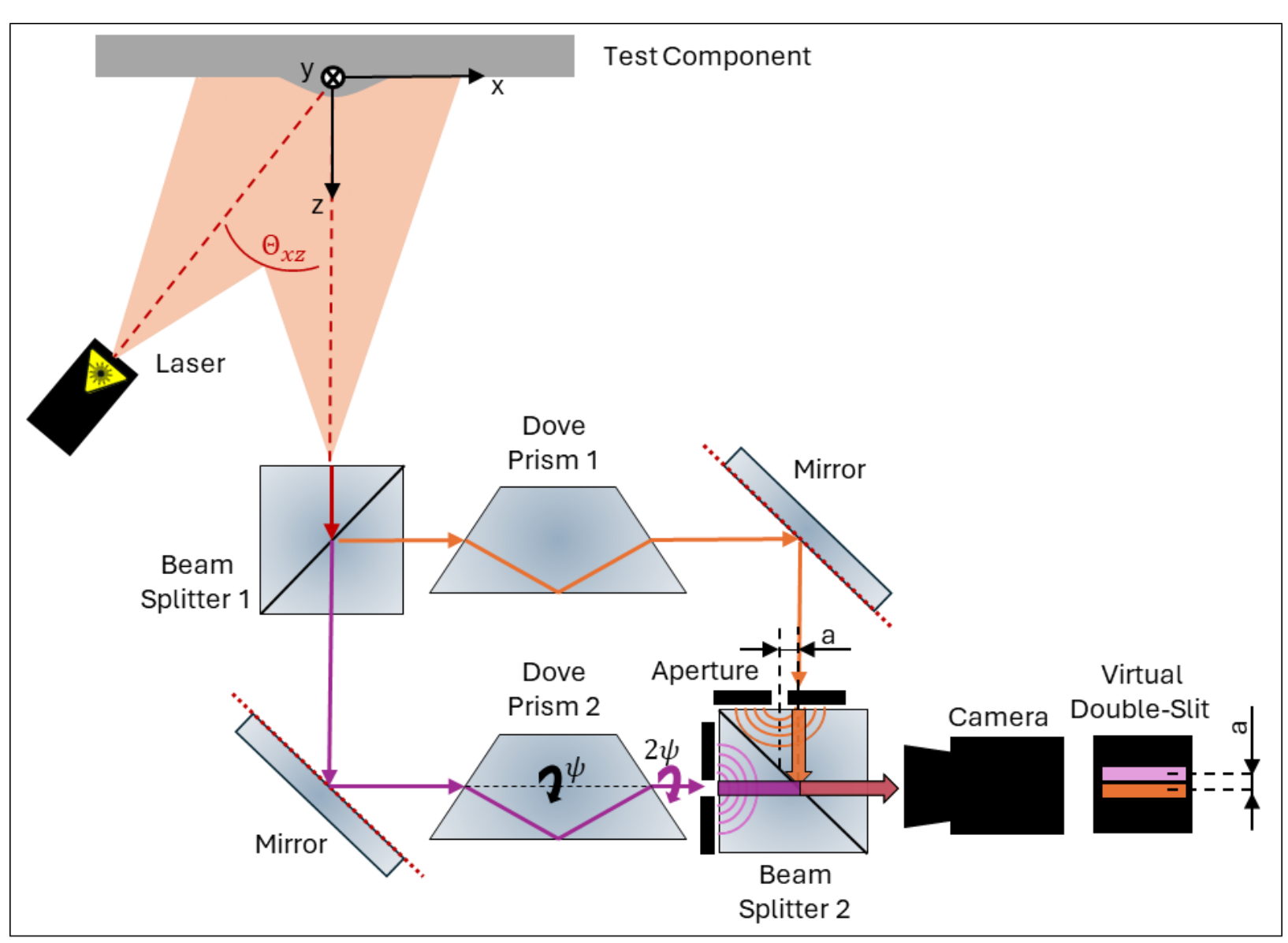

Figure 1: Mach-Zehnder interferometer configuration employing rotational shear. Rotation of one Dove prism by an angle $\psi$ results in an image rotation of $2\psi$. Owing to the off-axis arrangement of the two slit apertures at the camera-side beam splitter (BS2), a coherent intensity modulation is generated, which decouples the shear amount from the formation of the spatial carrier frequency.

For the following considerations, the VDS can be regarded as functionally equivalent to a conventional double slit, in which one wavefront (orange and violet, respectively) passes through each slit. At the two slit apertures, the incident wavefronts undergo diffraction according to the principle of Christiaan Huygens. Owing to the off-axis arrangement of one of the two slit apertures, an additional coherent intensity modulation is superimposed onto the two overlapping speckle patterns. This effect is schematically illustrated in figure 2.

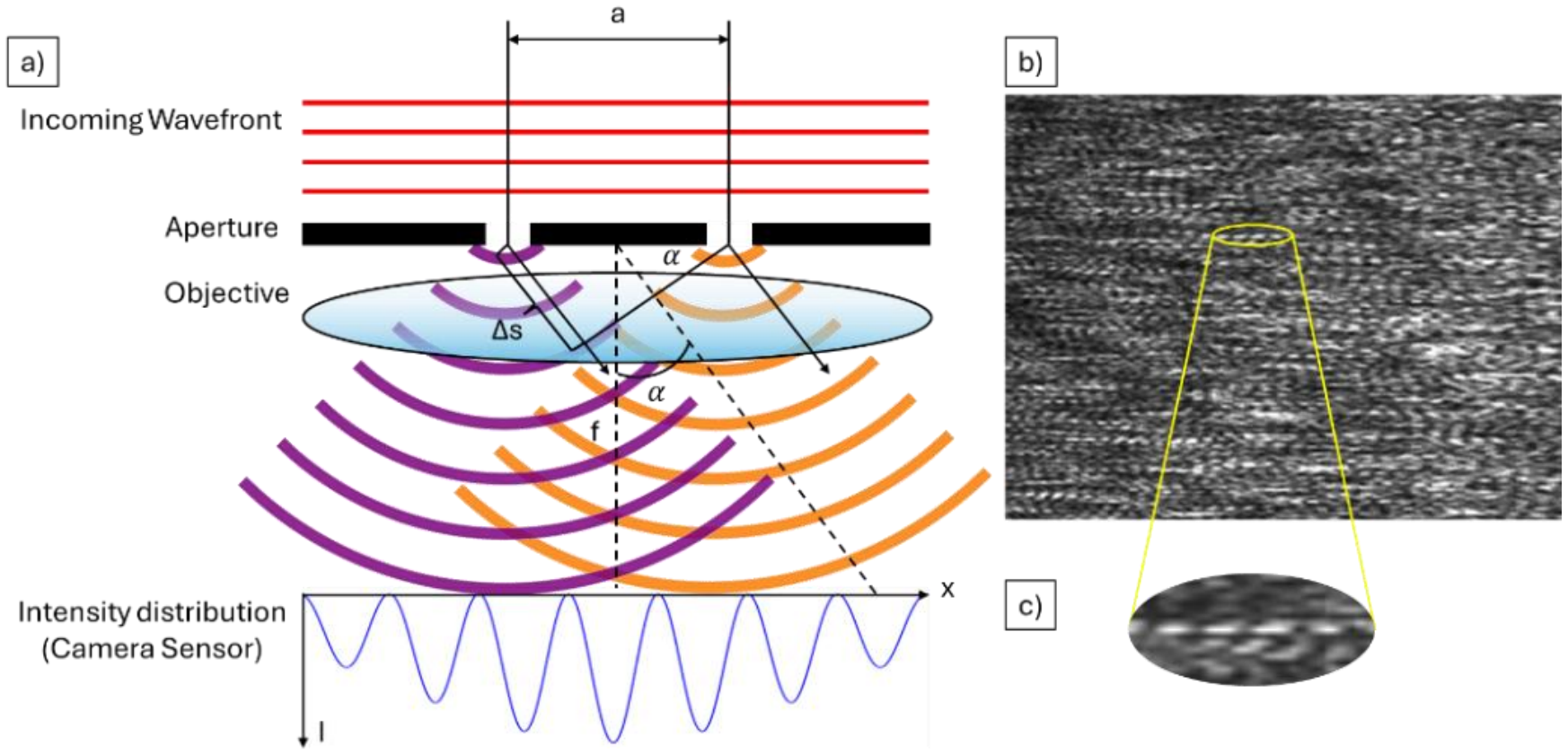

Figure 2: Interference after transmission through a VDS. a) At the slits, two new wavefronts are generated, whose wave fields superimpose. On an image sensor positioned at a distance equal to the lens focal length f from the double slit, an intensity distribution is formed, which is experienced only by the coherent laser light (adapted from [7], p.100). b) Live camera image of a speckle pattern. Due to the off-axis displacement of one of the two slit apertures within the optical setup, a coherent intensity modulation is introduced, resulting in elongated, modulated speckles. c) Magnified section of a modulated speckle.

The incident wavefront (shown in red in figure 2) is diffracted at the aperture, giving rise to new wavefronts. Due to the resulting optical path difference, these wavefronts undergo interferometric superposition in the image plane (camera sensor). The optical path difference $\Delta s$ is calculated according to the following equation [7], pp. 99 ff.:

$$\frac{\Delta s}{a} = \frac{x}{f} \tag{2}$$

| | |
|---|---|
| $\Delta s$ | Optical path difference |
| $a$ | Slit spacing |
| $x$ | Spatial coordinate |
| $f$ | Focal length |

The corresponding phase difference is obtained from the optical path difference as follows:

$$\frac{\Delta s}{\lambda} 2\pi = \phi_0 \tag{3}$$

| | |
|---|---|
| $\lambda$ | Laser wavelength |
| $\phi_0$ | Phase difference |

For signal processing, the methodology according to Takeda [9] is employed. Figure 3 illustrates the influence of different slit-aperture arrangements within the optical setup on the spatial frequency spectrum under rotational shear. The spatial carrier-frequency spectra relevant for the measurement are highlighted in green, while the spectrum associated with the background light intensity is marked in red.

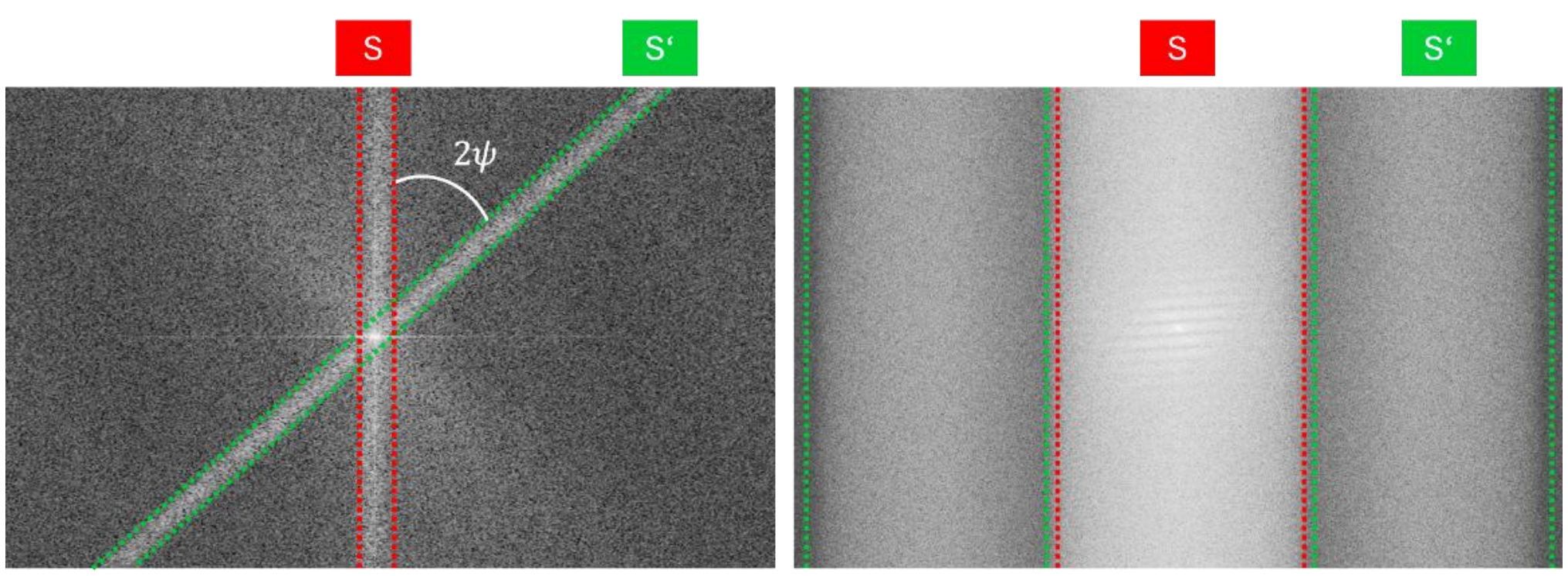

Figure 3: Influence of different slit-aperture arrangements within the optical setup for a rotational shear of 20°. Left: A conventional SPS slit-aperture arrangement positioned in front of the optical setup causes the spatial carrier-frequency spectra to rotate along with the image rotation angle. Right: The use of a VDS within the optical setup (see Fig. 1) results in a spatial carrier-frequency formation that is decoupled from the adjustment of the shear amount.

In the right-hand image, a VDS is employed within the optical setup to generate the spatial carrier frequencies. In contrast, the left-hand image uses a single-slit aperture (slit width approximately 0.1 mm) positioned in front of the optical setup. Both amplitude spectra are observed when one Dove prism in the optical system is rotated by $\psi = 10°$, resulting in an image rotation of $2\psi = 20°$. When a preceding slit aperture is used (left-hand image in fig. 3), the entire spectrum S‘ (highlighted in green) rotates together with the image by 20°. In contrast, the use of a VDS ensures a constant formation of the spatial carrier-frequency spectra independent of the prism rotation. This allows SPS-Shearography to be applied in a manner decoupled from the shear amount, thereby enabling flexible sensitivity adjustment for different measurement tasks. From Eqs. (2) and (3), the expression for calculating the desired carrier phase $\phi_0$ follows as:

$$\phi_0 = 2\pi \frac{a}{\lambda f} x = 2\pi f_0 x \tag{4}$$

$f_0$ Spatial carrier frequency

For an interferometric image acquisition containing a spatial carrier frequency, the carrier phase is added to the phase term in the intensity equation, resulting in the following expression:

$$I = I_0(1 + \gamma \cos(\phi + \phi_0)) = I_0(1 + \gamma \cos(\phi + 2\pi f_0 x)) \tag{5}$$

For a given wavelength and lens focal length, the slit width and slit separation of the VDS can be designed to establish an optimal spatial carrier frequency $2\pi f_0 x$. This enables the use of a constant FFT amplitude spectrum for spatial phase shifting, decoupled from the applied shear amount [7], pp. 101 ff.

## 3. Measurement Principle using rotational Shear

To obtain shearograms containing interference fringes, the components under investigation must be examined under different loading states. The initial or reference state is defined as the condition of the test object at which the first reference image is acquired [10], p. 16. This state may correspond to either a loaded or an unloaded condition. It is essential that the subsequent loading state differs from the reference state in terms of the applied load magnitude. As a consequence of the load difference applied to the component under investigation, a deformation difference arises. As a result, an individual object point $P$ on the surface of the component undergoes a small displacement. A neighbouring point $Q$ (arising from the double image caused by shearing) undergoes a slightly different displacement, leading to a differential displacement between the two points. This situation is examined in more detail with the aid of fig. 4. Here, $x_s$, $y_s$, $z_s$ denote the coordinates of the laser light source, while $x_o$, $y_o$, $z_o$ represent those of the observation plane. The distances between point P and the laser light source $S$ and between point $P$ and the observation plane $O$ are denoted by $R_s$ and $R_o$ respectively.

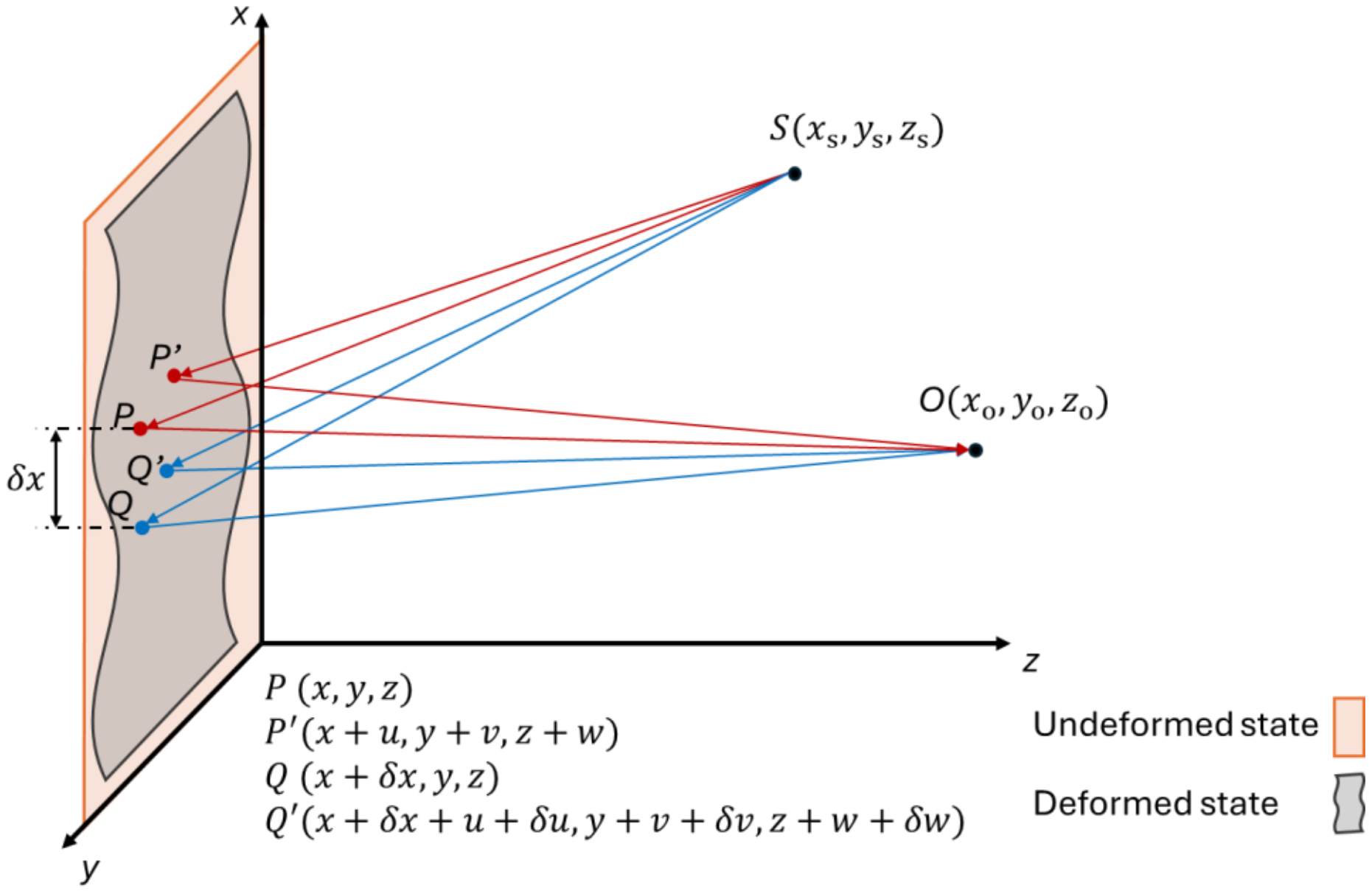

Figure 4: Illustration of the optical path difference of the laser light between two sheared points on the surface of the test object in the undeformed and deformed state (adapted from [11])

The relative phase change $\Delta$ is induced by the change in the relative optical path length of light scattered from two neighbouring object points. For conventional linear shearing direction, the distance between the neighbouring points $P(x, y, z)$ and $Q(x + \delta x, y, z)$ is the amount of shearing $\delta x$ (the same applies to shearing in the $y$-direction.) The relative phase change $\Delta$ is given by the following expression [10], p. 16 f.:

$$\Delta = \frac{2\pi}{\lambda}(\delta L_1 - \delta L_2) = \frac{2\pi}{\lambda}(A\delta u + B\delta v + C\delta w) \tag{6}$$

where $$A = \left(\frac{x - x_o}{R_o} + \frac{x - x_s}{R_s}\right); B = \left(\frac{y - y_o}{R_o} + \frac{y - y_s}{R_s}\right); C = \left(\frac{z - z_o}{R_o} + \frac{z - z_s}{R_s}\right) \tag{7}$$

with $$\delta u = u(Q) - u(P)\,; \delta v = v(Q) - v(P)\,; \delta w = w(Q) - w(P) \tag{8}$$

Where $\delta L_1$ represents the change in the optical path length of the unsheared point $P$ and $\delta L_2$ the change in the optical path length of the sheared point $Q$, both measured from the illumination source to the observation plane. Furthermore, $R_o$ represents the distance between the observation plane $O$ and point $P$ and $R_s$ the distance between the laser light source $S$ and point $P$. The quantities $u$, $v$ and $w$ denote the displacement components of an object point between the reference and the loaded state in the $x$-, $y$- and $z$-directions, respectively. The quantities $\delta u$, $\delta v$ and $\delta w$ represent the relative displacement differences between points $P$ and $Q$ comparing the reference and the loaded state. For small shear amounts, the displacement difference between two neighbouring object points can be approximated by the first-order term of a Taylor expansion of the displacement field [12]. For an arbitrary displacement component $f \in \{u, v, w\}$, this gives:

$$f(Q) - f(P) = \nabla f(P) \cdot (Q - P) + \mathcal{O}(|\,Q - P\,|^2) \tag{9}$$

$$f(Q) - f(P) \approx \frac{\partial f}{\partial x}(x_Q - x) + \frac{\partial f}{\partial y}(y_Q - y) + \frac{\partial f}{\partial z}(z_Q - z) \tag{10}$$

Equation (10) represents the Cartesian component form of the first-order Taylor approximation given in Equation (9). The type of shearing amount determines the local shear vector $Q - P$. For linear shearing in the $x$-direction, the sheared point is $Q = (x + \delta x, y, z)$ and therefore:

$$x_Q - x = \delta x\,; y_Q - y = 0\,; z_Q - z = 0 \tag{11}$$

Using Eq. (9), the displacement differences for linear shearing in $x$-direction become:

$$\delta u \approx \frac{\partial u}{\partial x}\delta x\,; \delta v \approx \frac{\partial v}{\partial x}\delta x\,;\ \delta w \approx \frac{\partial w}{\partial x}\delta x \tag{12}$$

Substitution into Eq. (6) gives:

$$\Delta_x = \frac{2\pi}{\lambda}\left(A\frac{\partial u}{\partial x} + B\frac{\partial v}{\partial x} + C\frac{\partial w}{\partial x}\right)\delta x \tag{13}$$

Analogously, for linear shearing in the $y$-direction, $Q = (x, y + \delta y, z)$, and thus:

$$x_Q - x = 0\,; y_Q - y = \delta y\,; z_Q - z = 0 \tag{14}$$

This leads to:

$$\Delta_y = \frac{2\pi}{\lambda}\left(A\frac{\partial u}{\partial y} + B\frac{\partial v}{\partial y} + C\frac{\partial w}{\partial y}\right)\delta y \tag{15}$$

In Equations (13) and (15), $\delta x$ and $\delta y$ denote the linear shear amounts in the $x$- and $y$-direction, respectively. Using rotational shearing, Point $P$ is characterized by a polar angle $\psi$ measured with respect to the image centre. When a rotational shear amount is introduced, the sheared neighbouring point $Q$ results from an additional rotation by the shear angle $\delta\psi_\mathrm{s}$. The rotation takes place within the *xy*-plane, while the *z*-coordinate remains unchanged. Therefore, the position of point $P$ is described using planar polar coordinated as:

$$x = r\cos\psi\,;\, y = r\sin\psi \tag{16}$$

A rotation by the shear angle $\delta\psi_\mathrm{s}$ changes the polar angle $\psi$ to $\psi + \delta\psi_\mathrm{s}$, while the radius $r$ remains constant. Therefore, the coordinates of the rotationally sheared point $Q$ are:

$$x_Q = r\cos\left(\psi + \delta\psi_s\right);\; y_Q = r\sin\left(\psi + \delta\psi_s\right) \tag{17}$$

By applying the trigonometric addition theorems, this gives:

$$x_\mathrm{Q} = r(\cos\psi\cos\delta\psi_\mathrm{s} - \sin\psi\sin\delta\psi_\mathrm{s}) \tag{18}$$
$$y_\mathrm{Q} = r(\sin\psi\cos\delta\psi_\mathrm{s} + \cos\psi\sin\delta\psi_\mathrm{s}) \tag{19}$$

By substituting $x = r\cos\psi$ and $y = r\sin\psi$, the following expressions are obtained:

$$x_\mathrm{Q} = x\cos\delta\psi_\mathrm{s} - y\sin\delta\psi_\mathrm{s} \tag{20}$$
$$y_\mathrm{Q} = x\sin\delta\psi_\mathrm{s} + y\cos\delta\psi_\mathrm{s} \tag{21}$$
$$z_\mathrm{Q} = z \tag{22}$$

Thus, the local shear vector components $Q - P$ are:

$$x_\mathrm{Q} - x = x(\cos\delta\psi_\mathrm{s} - 1) - y\sin\delta\psi_\mathrm{s} \tag{23}$$
$$y_\mathrm{Q} - y = x\sin\delta\psi_\mathrm{s} + y(\cos\delta\psi_\mathrm{s} - 1) \tag{24}$$
$$z_\mathrm{Q} - z = 0 \tag{25}$$

For small shear amounts, the trigonometric functions can be approximated by their first-order expressions:

$$\sin\delta\psi_\mathrm{s} \approx \delta\psi_\mathrm{s} \tag{26}$$
$$\cos\delta\psi_\mathrm{s} \approx 1 \tag{27}$$

Using these approximations in Eqs. (23) and (24), the local rotational shear vector becomes:

$$x_\mathrm{Q} - x \approx -y\delta\psi_\mathrm{s} \tag{28}$$
$$y_\mathrm{Q} - y \approx x\delta\psi_\mathrm{s} \tag{29}$$
$$z_\mathrm{Q} - z = 0 \tag{30}$$

Using the first-order approximation from Eq. (10) and substituting Eqs. (28) – (30), this gives

$$f(Q) - f(P) \approx \left(-y\frac{\partial f}{\partial x} + x\frac{\partial f}{\partial y}\right)\delta\psi_\mathrm{s} \tag{31}$$

Accordingly, the relative displacement differences of the three displacement components are obtained as:

$$\delta u \approx \left(-y\frac{\partial u}{\partial x} + x\frac{\partial u}{\partial y}\right)\delta\psi_{\mathrm{s}} \tag{32}$$

$$\delta v \approx \left(-y\frac{\partial v}{\partial x} + x\frac{\partial v}{\partial y}\right)\delta\psi_{\mathrm{s}} \tag{33}$$

$$\delta w \approx \left(-y\frac{\partial w}{\partial x} + x\frac{\partial w}{\partial y}\right)\delta\psi_{\mathrm{s}} \tag{34}$$

Substituting Eqs. (32) – (34) into the general phase relation in Eq. (6) gives the relative phase change when rotational shear is applied:

$$\Delta_{\mathrm{rot}} \approx \frac{2\pi}{\lambda}\left[A\left(-y\frac{\partial u}{\partial x} + x\frac{\partial u}{\partial y}\right) + B\left(-y\frac{\partial v}{\partial x} + x\frac{\partial v}{\partial y}\right) + C\left(-y\frac{\partial w}{\partial x} + x\frac{\partial w}{\partial y}\right)\right]\delta\psi_{\mathrm{s}} \tag{35}$$

For linear shearing in the $x$- or $y$-direction, the shear vector is given by $Q - P = (\delta x, 0,0)$ or $Q - P = (0, \delta y, 0)$, respectively (see Eqs. (11) and (14)). Therefore, its magnitude is constant over the entire image and equal to $\delta x$ or $\delta y$. If rotational shearing is applied, the local shear vector depends on the position of the object point relative to the centre of rotation. Using Eqs. (30) and (31), the magnitude of the local rotational shear vector is obtained as:

$$\mid Q - P \mid \approx \sqrt{(-y\delta\psi_{\mathrm{s}})^2 + (x\delta\psi_{\mathrm{s}})^2} \tag{36}$$

$$\mid Q - P \mid \approx \delta\psi_{\mathrm{s}}\sqrt{x^2 + y^2} \tag{37}$$

With $r = \sqrt{x^2 + y^2}$ this becomes

$$\mid Q - P \mid \approx r\delta\psi_{\mathrm{s}} \tag{38}$$

The expression $-y\frac{\partial f}{\partial x} + x\frac{\partial f}{\partial y}$ corresponds to the derivative of the displacement component $f$ with respect to the angle $\psi$ at constant radius. Thus, the phase change for rotational shear can be also written as:

$$\Delta_{\mathrm{rot}} \approx \frac{2\pi}{\lambda}\left(A\frac{\partial u}{\partial \psi} + B\frac{\partial v}{\partial \psi} + C\frac{\partial w}{\partial \psi}\right)\delta\psi_{\mathrm{s}} \tag{39}$$

This shows that rotational shearography is sensitive to tangential (azimuthal) displacement gradients around the centre of rotation. Using rotational shearing, the local shear amount increases linearly with the distance $r$ from the centre of rotation. It is zero at the image centre and becomes larger towards the outer regions of the image. Consequently, rotational shearing produces a position-dependent shear amount, whereas linear shearing produces a constant shear amount over the image field.

## 4. Experimental Validation

For experimental validation, a test setup is presented in which a calibration specimen is investigated using SPS-Shearography under pneumatic overpressure excitation. The newly developed shearographic system employing rotational shear is utilized. The illumination angle is adjusted such that a pure out-of-plane (OOP) measurement sensitivity is achieved. The test specimen consists of an aluminium body containing circular artificial defects (flatbottom holes) of varying diameters. These defects were introduced using an end mill. The prepared flaws exhibit different residual wall thicknesses, as summarized in Table 1:

| Designation | Defect Diameter [mm] | Residual wall thickness [mm] |
|---|---|---|
| A | 18 | 2 |
| B | 10 | 2 |
| C | 12 | 3 |
| D | 16 | 3 |
| E | 18 | 3 |
| F | 10 | 1 |
| G | 12 | 1 |
| H | 16 | 1 |

Table 1: Defect parameters and corresponding residual wall thicknesses

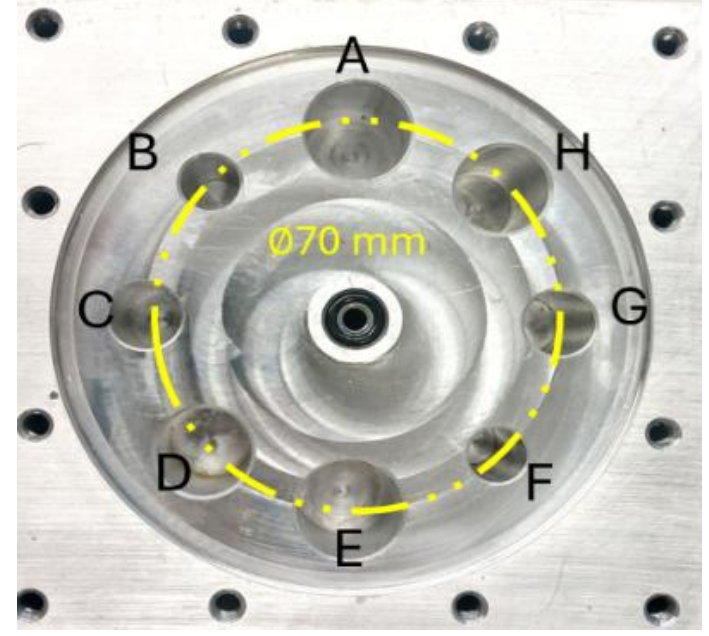

Figure 5: Back side of the test specimen with diameter designation

The excitation is performed using internal overpressure. Sealing is achieved by installing a backing plate with an nitrile butadiene rubber sealing ring on the rear side. The defects are arranged on a circular arc with a diameter of 70 mm.

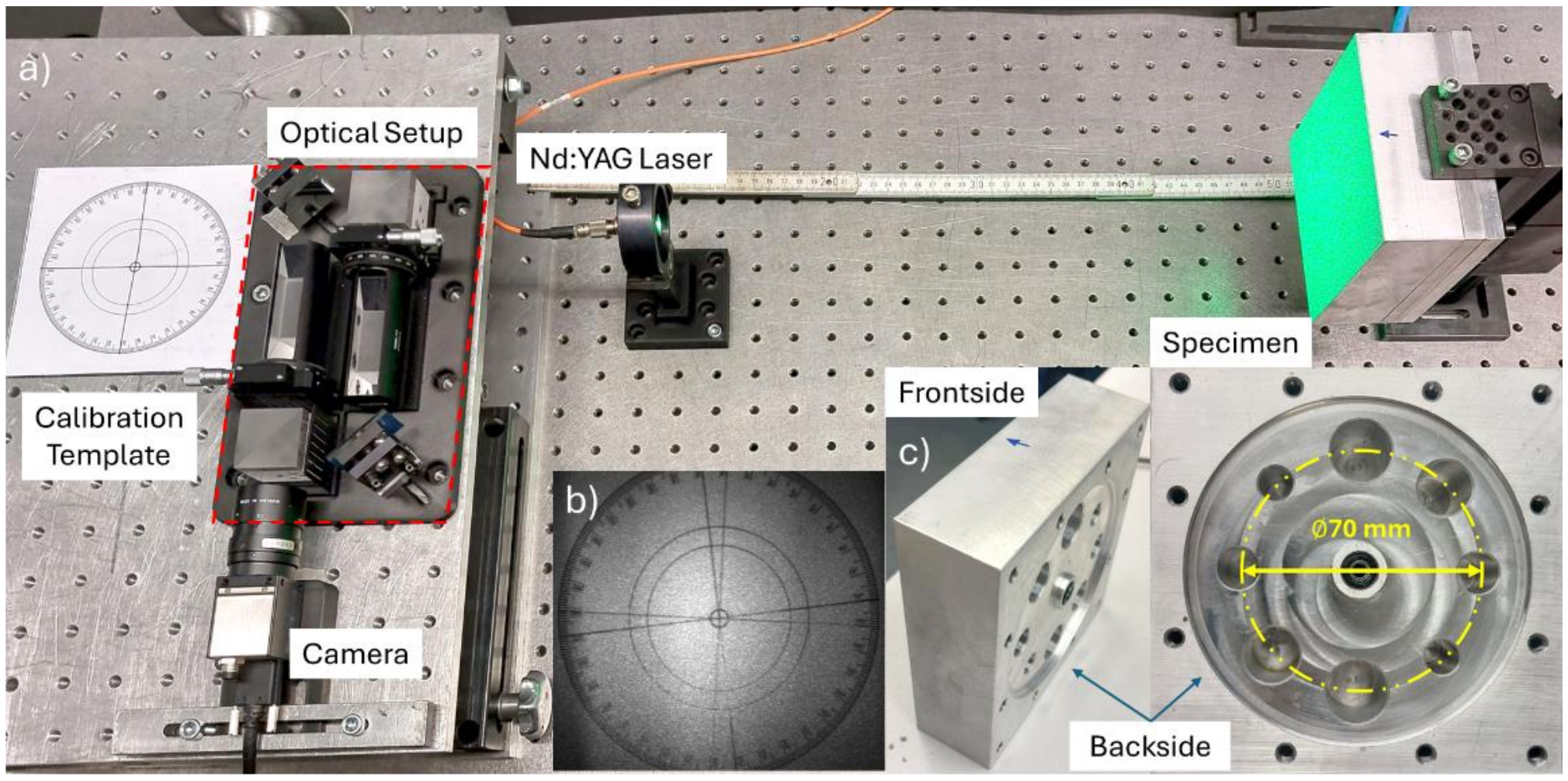

Figure 6: a) Experimental setup for the investigation of an overpressure-loaded test specimen featuring eight flat-bottom holes with varying residual wall thicknesses, arranged at equal angular intervals (45°) on a pitch circle with a diameter of 70 mm, b) Live camera image of the calibration template under rotational shear of 10°, c) View of the opened specimen showing circular defects located on the pitch circle.

Figure 6 illustrates the experimental setup and the calibration target. The specimen is illuminated using an expanded Nd:YAG laser beam with a wavelength of 532 nm at a measurement distance of approximately 500 mm. The target is examined in two loading states. In the loaded condition, a pressure increase of $\Delta p = 3$ bar is applied relative to the reference state. The following figure presents representative shearographic measurement results. During the experiment, four different shear configurations are investigated:

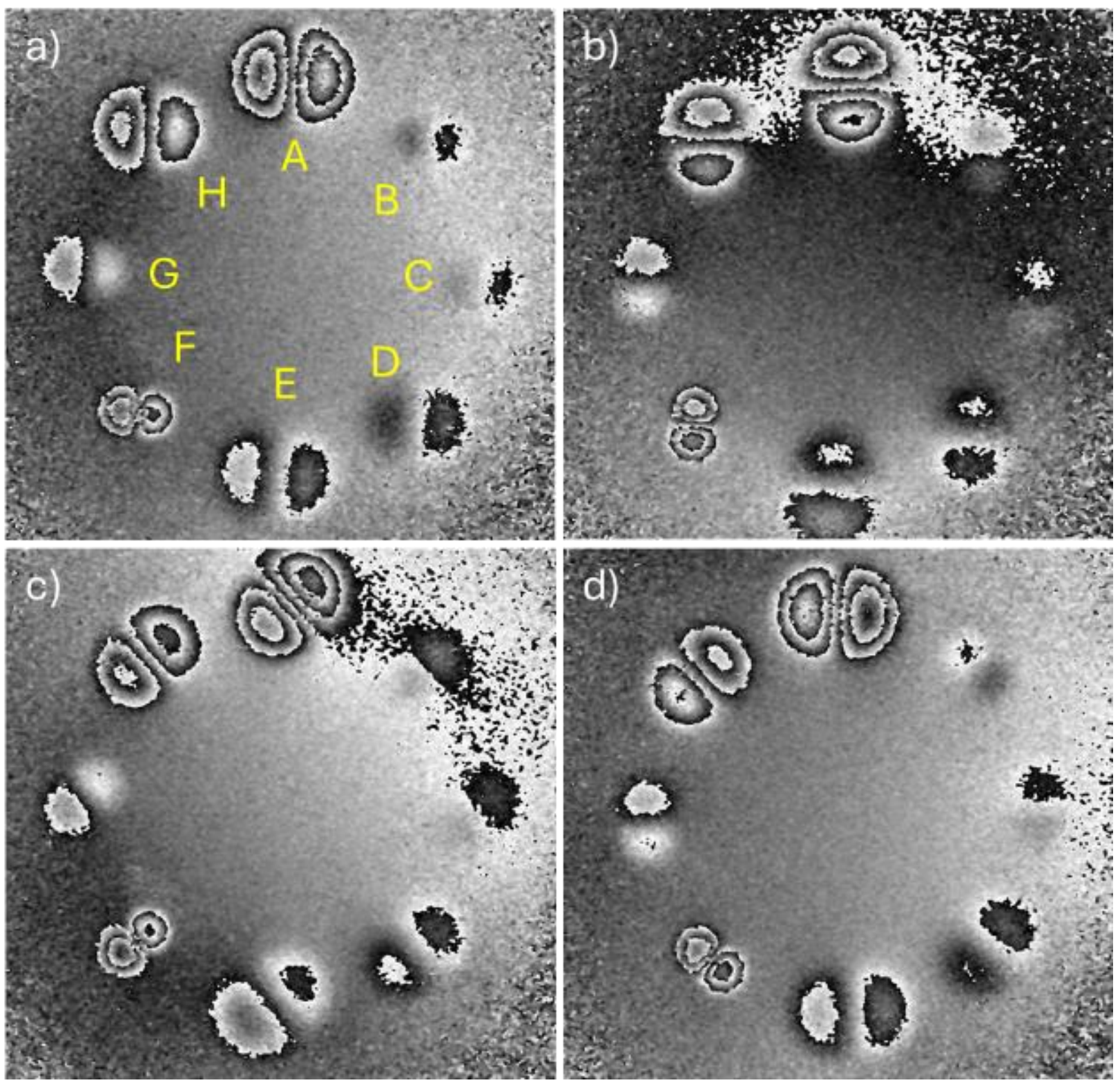

Figure 7: Shearographic measurement results of an overpressure-loaded specimen under four different shear configurations. a) 10 mm shear in *x*-direction with highlighted defects (yellow), b) 10 mm shear in *y*-direction, c) 10 mm shear in *xy*-direction, d) rotational shear of approximately 10°.

The conducted measurements demonstrate that all defects, irrespective of their diameter and residual wall thickness, can be reliably detected using both conventionally applied linear shear configurations (Fig. 7(a) – (c)) as well as rotational shear. The circular defects appear in the measurement results as characteristic shearographic butterfly patterns. A distinctive feature of the measurements obtained with rotational shear is the orientation of the deformation gradient in the tangential (azimuthal) direction around the image centre. This implies that, when inspecting actual samples, there is increased sensitivity to defects such as cracks that propagate radially (from the center of the image toward the outer diameter).

## 5. Conclusion

This work presents the development and experimental validation of a novel shearographic setup for spatial phase-shifting (SPS) shearography employing rotational shear. The proposed optical configuration with VDS enables adjustment of the shearing amount independently of the spatial carrier frequency, thereby allowing unrestricted sensitivity tuning. Experimental investigations on a pneumatically loaded test specimen with circular defects of varying residual wall thickness demonstrate that rotational shear can be reliably applied for defect detection. The measurement system exhibits pronounced sensitivity to deformation gradients in the tangential direction around the image centre, highlighting its suitability for detecting defects that propagate radially. Future work will focus on further developing the presented setup to enable measurements under industrial conditions, as well as investigating additional test specimens and application scenarios. The use of SPS-shearography opens up a wide range of possibilities for the inspection of different objects in real industrial environments. This approach is of particular relevance for safety-critical sealing components, for example in the aerospace, nuclear, or oil and gas sectors[13]. Furthermore, the proposed approach shows high potential for in situ monitoring of sealing rings and may extend existing research efforts [14]. According to the current state of the art, rotationally symmetric components (especially thin-walled sealing rings) can in principle be examined under industrial conditions, using SPS-shearography. However, it is not possible to examine the entire surface of thin-walled specimens using a single sensor or with a single specimen orientation. By adjusting a linear shearing amount, the usable measurement area becomes highly limited for rotationally symmetric components with thin walls (see fig. 8 a), limited measuring areas are marked in red). As a result, multiple measurements are required, with either the specimen or the sensor being rotated stepwise depending on the component diameter in order to inspect the entire surface using shearographic measurement. In contrast, the newly developed method employing rotational shearing amount leads to a significantly faster inspection time for full-field evaluation (see fig. 8 b), where the unlimited measuring area is highlighted in green).

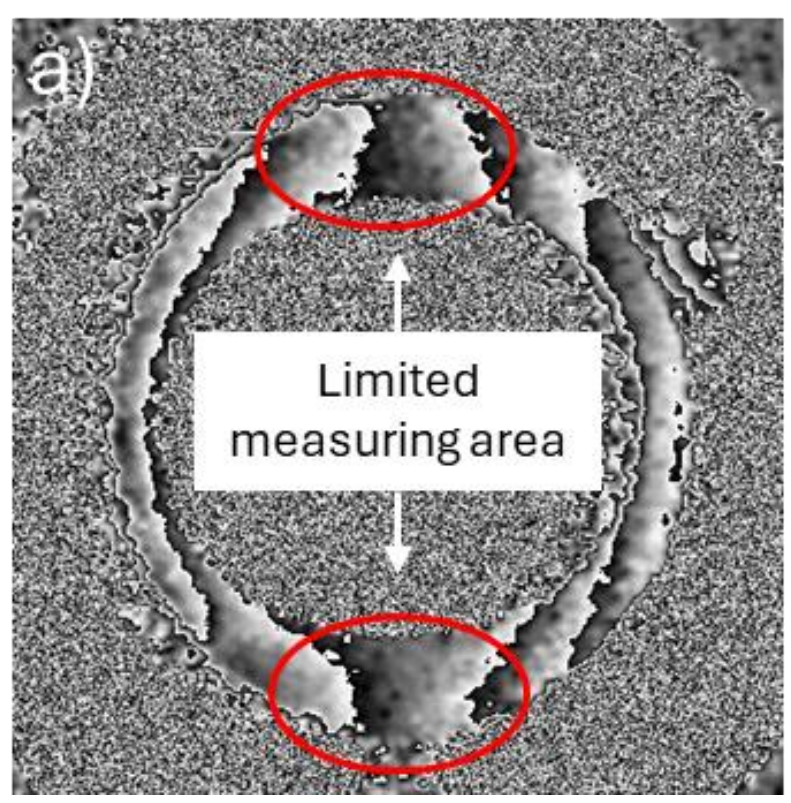

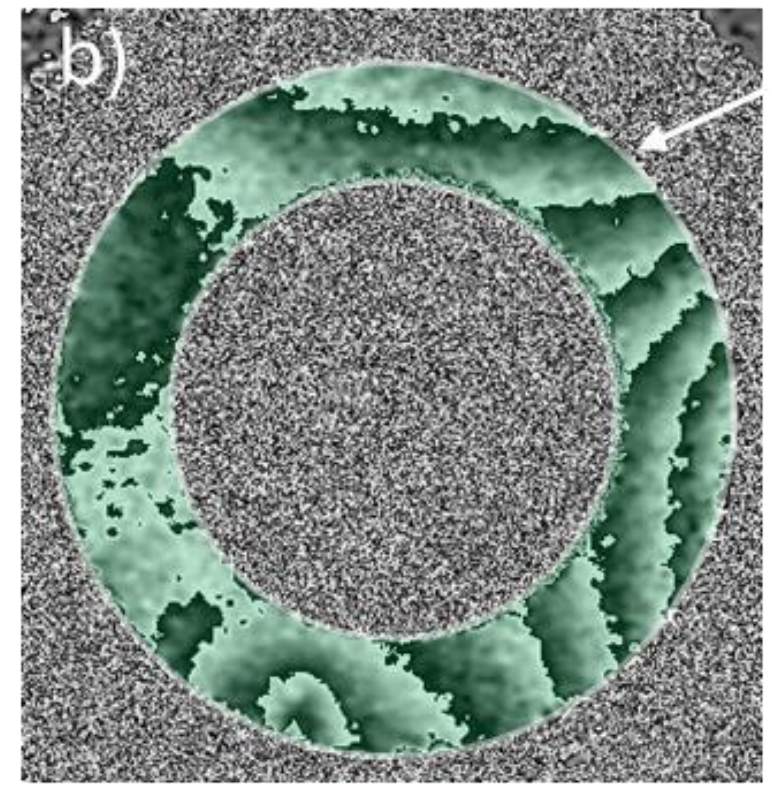

Figure 8: Non-destructive testing of a radial shaft seal using thermally excited SPS-shearography, presented as an application example. Left: Linear shear in the *x*-direction, resulting in a limited measurement area and requiring multiple rotation of the specimen or sensor for full inspection. Right: Rotational shear enabling full-field measurement coverage, allowing a fast one-shot acquisition without the need for repositioning.

A characteristic aspect of the rotational shear approach that should be considered, particularly in the inspection of sealing rings, is its reduced sensitivity to defects oriented in the circumferential direction (e.g., cracks that are propagated in tangential direction), as these align with the shear direction and therefore exhibit lower contrast in the measured deformation gradient. In such cases, radial shear configurations may provide enhanced sensitivity and should therefore be further investigated. Moreover, a combination of rotational and radial shear appears promising and warrants further study. Furthermore, future work will also focus on extending the proposed rotational shear approach from defect detection toward defect depth characterization [15].

**Funding.** This work was funded by the Joint Science Conference (GWK), the Federal Ministry of Education and Research (BMBF), and the state of Rhineland-Palatinate.

**Acknowledgment.** The Authors acknowledge the Joint Science Conference (GWK), the Federal Ministry of Education and Research (BMBF), and the state of Rhineland-Palatinate for funding. We would also like to thank the German Society for Non-Destructive Testing (DGZfP) for supporting Mr. Bastgen through the DGZfP scholarship.

**Disclosures.** The authors declare no conflicts of interest.

**Data availability.** Data underlying the results presented in this paper are not publicly available at this time but may be obtained from the authors upon reasonable request.